\documentstyle[preprint,aps,epsfig] {revtex}
\tighten
\begin{document}

\title{Fixed energy inversion of 5 eV e-Xe atom scattering}
\author{A. Lovell and K. Amos}
\address{ School of Physics, University of Melbourne,\\
Victoria 3052, Australia.}
\date{\today}
\maketitle
\begin{abstract}
Fixed energy inverse scattering theory has been used 
to define central and spin-orbit Schr\"odinger potentials
for the scattering of
5 eV polarized electrons
 from Xe atoms. 
The results are typical for a range of such data;
including energies above threshold when the 
potentials become complex. 
The phase shifts obtained from an analysis
of the measured differential cross section and analyzing power
has been used as input data.
Both semi-classical (WKB) and fully quantal
inversion methods have been used to extract central and spin-orbit
interactions.
The analysis shows that information
additional to the set of input phase shifts 
extracted from this (and similar) data
may be
needed to ascertain  physical potentials.
\end{abstract}


\section{Introduction}

A knowledge of the interaction between
colliding quantum systems 
is central in many applications 
of scattering and 
has relevance for use in 
other diverse fields 
of study.
Such
interactions have been sought in various ways with 
the method of numerical inversion 
common. In the numerical inversion  approach,
the parameter values of a
purely phenomenological parametric form, chosen 
{\it a priori} to be the 
(central, local) interaction
between the colliding entities, are
determined by variation until a best fit 
to measured data is found.
Global inverse scattering methods~\cite{C&S} 
form an alternative procedural class  with
which to analyze the same data.  With 
global inverse scattering methods, the interaction
between the colliding pair is extracted from the data
without {\it a priori} assumptions about the shape of the potential,
although it may belong to a certain broad class, and the validity
of the  dynamical equation of motion (the Schr\"{o}dinger equation) is assumed.
Potentials so obtained we define hereafter as inversion potentials.
Application of various global inverse scattering methods have
been made in the past for electron-atom~\cite{AlleneA},
for atom-atom~\cite{AtomAtom}, and for electron-molecule~\cite{electwater}
systems, but none of those methods permitted extraction of
spin-orbit effects.  Recent developments~\cite{LHF,Leeb,Lun}
have provided means by which spin effects can be treated.
In this paper we present and describe results for central and spin-orbit
potentials obtained by global inverse scattering methods.
In particular we consider an approach based upon
the WKB approximation method~\cite{C&S,WKB}
and one based upon 
the Newton-Sabatier (N-S)        
theory~\cite{C&S}.

The data of interest come from the
very high quality crossed
beams experiment of Gibson {\it et al.}~\cite{Buckman}.
In that study~\cite{Buckman}, a phase shift analysis was also
made and the phase shifts so obtained
were the input
quantities to our studies.  Those phase shifts are purely real in line with
the unitarity constraint with the energy below the first threshold.
Thus we have obtained purely real inversion potentials.
Extension of the approach to deal with energies above threshold
and, concomitantly with complex potentials, is straightforward. 
The key feature about inversion potentials, given numerical
accuracy in calculations and stability of the solution,
is that when used in Schr\"odinger equations they
lead to the same phase shifts as are input.

The data set we have chosen to use
is intriguing for a number of reasons. 
First, the differences between the
values of $\delta^+_l$ and $\delta^-_l$
(the $\pm$ superscripts denoting $j = l \pm \frac{1}{2}$)
 extracted by
the phase shift analysis~\cite{Buckman} are 
not large indicating that the spin-orbit
interaction is not strong. As a result the 
approximation method of Leeb,
Huber and Fiedeldey~\cite{LHF}
(LHF hereafter)  can be used 
with confidence. 
Indeed the LHF scheme 
is accurate through second order in Born approximation
and has worked well in some nuclear scattering
data analyses~\cite{LHFnuc}
where the spin-orbit effect is much stronger than in the case
we study.
With the LHF approximation, both 
the N-S and the semi-classical WKB 
methods of inverse scattering
theory can be used  
to specify the electron-Xenon (e-Xe) potentials. 
Exact quantal inversion  methods to get the spin-orbit
interaction are known~\cite{Leeb,Lun},
but with this data the LHF approximation should be adequate
and the inversion process is facilitated by its use.
Second,
the phase shifts of significance
 are not many in number and so this may be 
another case with which phase shift
values at unphysical rational values of angular momentum are 
required in the inversion process
to achieve a stable result~\cite{MEpaper}. 
A third reason for interest is that the $s$- and $p$-wave
 phase shifts
from the analysis~\cite{Buckman} 
of the scattering data have negative values.
All other (physical) phase shifts of significance are positive
quantities. As phase shifts
are ambiguous to modulo $\pi$, an equivalent completely 
positive valued and monotonically
decreasing phase shift set can be formed by the addition 
of $2\pi$ and $\pi$ to the $s$- and $p$-wave values
respectively. With either of the original or
the modulated
(integer $l$) phase shift 
sets as input, 
 the N-S inversion method {\it per se} 
gives the same inversion potential. 
However, that potential has
a short ranged repulsion; such being required~\cite{Newton}
 to give significant 
negative $s$- and $p$-wave phase shifts.
The WKB inversion method on the other hand, 
does discriminate between these sets since to specify
the WKB inversion potential interpolated functions, 
$\delta_l(\lambda)$, are required as input.
Of course if the N-S method is extended to use 
phase shifts at rational values of the angular momentum,
found for example by
interpolation of the original sets~\cite{Buckman}
and of the those with $s$- and $p$-wave values 
adjusted by $2\pi$ and $\pi$ respectively, 
then the inversion potentials from the two cases must differ.

Thus we give two new elements in this paper.  First we deduce
by inverse scattering theories, central
and spin-orbit potentials for 5 eV electron-Xenon atom scattering.
Second, we show that information
additional to the physical phase shifts (i.e. those
determined by usual phase shift analyses of scattering data)
is needed to identify the the most likely physical inversion potential.
In the next section we give a pr$\acute {\rm e}$cis of the 
LHF approximation for phase shifts as well as of the
inverse scattering methods used, the N-S 
and the WKB semi-classical schemes specifically.
Then in Sec.~III, we discuss the origins and characteristics of the
5 eV e-Xe scattering
phase shifts that have been used as
input to our inversion studies.
The e-Xe potentials that result are presented
 and discussed in Sec.~IV and we draw conclusions in Sec.~V.


\section{Fixed energy inversion methods}

In this section we give brief outlines of the methods used
in the calculations, the results of which we report later.
First we set out the LHF scheme by which the spin-orbit
interaction can be defined from two independent (spin-less)
inversion calculations.  This identifies not only the
special phase shift sets $\{ \hat \delta_l \}$
and $ \{ \tilde \delta_l \}$ of the method,
 but also defines the central and
spin-orbit potentials in terms of the results of inversions
of those new phase
shift sets, 
$\hat V$ and $\tilde V$ 
respectively.
Then we give the salient features of the Newton-Sabatier 
and the semi-classical WKB inverse scattering theories
which we have used to specify the 
($\hat V$ and $\tilde V$) 
potentials.

\subsection{The LHF approximation}

While exact quantal inverse scattering theories
that yield central and spin-orbit interactions
from input scattering phase shift sets exist~\cite{Leeb,Lun},
Leeb, Huber, and Fiedeldey~\cite{LHF} 
developed an approximation scheme 
to transform the input phase shift sets
so that more facile quantal inverse scattering
methods, such as the N-S
scheme~\cite{C&S} and the semi-classical WKB
approximation~\cite{WKB},
can be used
to give results from which central and spin-orbit
potentials can be extracted.
Those more facile schemes 
do not allow for an angular momentum dependence
in the intrinsic equation of motion,
such as given by a spin-orbit potential.
They are designed only to
provide central local potential functions.

The LHF method is based on the assumption that the
contribution of the spin-orbit potential to the phase shifts 
can be evaluated
using a Distorted Wave Born Approximation. 
This technique has been
formulated specifically for spin $\frac{1}{2}$ particles 
incident on spin zero
targets and is accurate to  second order in the 
Born expansion~\cite{LHF}.
The approximation  identifies first the
special phase shift sets $\{ \hat \delta_l \}$
and $ \{ \tilde \delta_l \}$, and then defines the central and
spin-orbit potentials in terms of the results,
 $\hat V$ and $\tilde V$ respectively,
from inversion of those new phase
shift sets.

For spin $\frac{1}{2}$ particles incident on a spin $0$ target,
and allowing central and spin-orbit Schr\"odinger potentials,
the scattering is defined by
reduced radial Schr\"{o}dinger equations
\begin{equation}
\frac{d^2}{d\rho^2} 
- \frac{l(l+1)}{\rho^2} + 1 - 
\frac{1}{E} \left[V_{cen}(\rho)
+ a^\pm_l V_{so}(\rho)\right]
\psi^\pm_l(\rho) = 0\ ,
\end{equation}
where, with $\rho = k r$,
\begin{equation}
a^\pm_l = \frac{1}{\hbar^2}
<{\mathbf s \cdot l}>\ = \left\{
\begin{array}{cl}
l      &\ \ j = l + \frac{1}{2}\\
-(l+1) &\ \ j = l - \frac{1}{2}
\end{array}
\right.\
\end{equation}

If the spin-orbit term is relatively weak, the
usual scattering  phase shifts, $\delta^\pm_l$,
can be expanded in powers of $a^\pm_l$.
Specifically~\cite{LHF}
\begin{equation}
\delta^\pm_l = 
\delta^{cen}_l 
+ a^\pm_l C^{(1)}_l(k) + (a^\pm_l)^2 C^{(2)}_l(k) + \dots .
\end{equation}
While the leading term, 
$\delta^{cen}_l$, 
is due solely to the central
component of the potential, $V_{cen}$, higher terms
must be considered to define the spin-orbit properties.
But the specific analytic forms of the coefficients, $C^{(n)}_l$,
do not have to be
known to extract the central and spin-orbit potential values. 
The LHF approximation is initiated by
considering combinations of 
$\delta^+$ and $\delta^-$ from which separate inversion
potentials
can be
estimated.
The relevant
combinations are
\begin{equation}
\tilde{\delta}_l = 
\frac{1}{2l+1} \left\{(l+1)\delta^+_l + l\delta^-_l\right\} =
\delta^{cen}_l + l(l+1) C^{(2)}_l + \dots ,
\label{tilde}
\end{equation}
and
\begin{equation}
\hat{\delta}_l = 
\frac{1}{2l+1} \left\{l \delta^+_l + (l+1)\delta^-_l\right\} =
\delta^{cen}_l - C^{(1)}_l + (l^2 + l +1)C^{(2)}_l + \dots
\label{hat}
\end{equation}

To first order in $a^\pm_l$, 
these new phase shifts and their inversion potentials,
$\tilde{V}$ and $\hat{V}$, are
\begin{eqnarray}
\tilde{\delta}_l = \delta^{cen}_l 
&\  \leftrightarrow \  & \tilde{V} \sim V_{cen}\\
\hat{\delta}_l = \delta^{cen}_l - C^{(1)}_l 
&\  \leftrightarrow \  & \hat{V} \sim V_{cen}
- {1\over2}V_{so}\ .
\end{eqnarray}
As these new sets of phase shifts can be inverted independently using
any of the conventional
techniques,
the central and spin-orbit components can be identified then by
\begin{eqnarray}
V_{cen}(r) & \approx & \tilde{V}(r) \\
V_{so}(r) & \approx & 2[\tilde{V}(r) - \hat{V}(r)]\ .
\label{Leebpot}
\end{eqnarray}

\subsection{The N-S method}

Since the Newton-Sabatier 
inverse scattering theory and applications
have been widely reported, 
only pertinent
points of the scheme are presented herein. 
A full treatment on the development
of this method is given elsewhere~\cite{C&S}.

The N-S method is one of the most successful of the fixed energy
inversion methods. 
Very recently, it has been
applied successfully to electron-helium atom scattering~\cite{harman} using
as input,
experimental phase shifts of Nesbet~\cite{Nesbet} at low
$\ell$-values and dipole polarization phase shifts at the higher
$\ell$-values. 
But it is known~\cite{Loeffel} that fixed energy inverse scattering
theory requires  the $S$ 
matrix 
(equivalently the phase shifts)
as a function of the angular momentum variable if
one is to define uniquely the scattering potential.
This equates to knowing the $S$ matrix exactly at all of the
(infinite) set of physical $l$-values as then the unit step in 
the quantum number is infinitesimal against the range.  
Most
studies of the fixed energy inverse scattering problem,
and notably those involving the N-S method, 
have been applied using only the values of
the $S$ matrix specified at a finite set ($l_{max}$)
 of physical angular momentum values.  
In cases where there are relatively few important
partial wave phase shift values to be used,
it may be necessary to 
extend~\cite{MEpaper} the usual N-S formulation
to include rational values of angular momenta
to form the matrix inherent in the N-S scheme~\cite{C&S}.

However it serves to consider for this section, just the 
integer values of angular momentum for which
the Schr\"{o}dinger equations take the form 
($\rho = kr$ being dimensionless)
\begin{equation}
D^u (\rho )\ \phi ^u_l (\rho ) = l(l+1)\ \phi ^u_l (\rho )
\end{equation}
where the operator
\begin{equation}
D^u (\rho ) = {\rho^2} \left[ {d^2\over {d\rho ^2}} + 1 - U(\rho ) \right]
\end{equation}
with $U(\rho) = V(\rho)/E_{cm}$,
where
$E_{cm} = \left(\hbar k\right)^2/\left(2\mu\right)$ and 
$ \rho = kr$.
The solutions are subject to boundary conditions
\begin{equation}
\phi _l^u (\rho)
\mathop{\longrightarrow}_{\rho \to \infty} 
 A_l\ \sin \left( \rho - \frac {1}{2} l\pi + \gamma _l \right)
\end{equation}
with $\gamma_l$ being the relevant phase shifts
to be taken as input quantities.  The N-S
method gives as 
output
\begin{equation}
U(\rho ) = U_0 (\rho ) - {2\over \rho} {d\over {d\rho }} {1\over \rho }
K(\rho ,\rho )
\end{equation}
wherein $U_0$ is a reference potential and $K(\rho ,\rho )$ is the 
Jost transformation kernel 
which can be written as  the infinite sum of solution function products,
\begin{equation}
K(\rho ,{\rho^\prime} ) =  \sum_l
 c_l\ \phi^u_l (\rho )\ \phi_l^{u_0} 
({\rho^\prime} )  
\end{equation}
The solution functions (to $D^u$ ) can be expressed by the 
Newton equations~\cite{C&S}
\begin{equation}
\phi^u_l (\rho ) = \phi^{u_0}_l (\rho ) - 
\sum_{l^\prime}
 c_{l^\prime}\ L_{ll^\prime} (\rho )\ \phi^u_{l^\prime} (\rho )
\label{eqn15}
\end{equation}
where
\begin{equation}
L_{ll^\prime}(\rho ) = \int_0^\rho \phi^{u_0}_l (\rho^\prime)
\phi^{u_0}_{l^\prime} ({\rho^\prime}) {1\over{\rho^\prime}^2} d{\rho^\prime} 
\end{equation}
These equations are of central importance. From them one can determine the 
unknown quantities, $A_l$ and $(c_l A_l)$, 
by matching asymptotically to the defined 
boundary condition solutions for $\rho \ge \rho_0$, $\rho_0$
being the value at which the unknown quantal
interactions are presumed to be vanishingly small.  
There is also a presumption that the solution functions of 
the reference potential are completely known so that the initiating L-matrices 
can be defined exactly. The reference solutions are obtained from 
\begin{equation}
D^{u_0} (\rho )\ \phi^{u_0}_l (\rho ) = l(l+1)\ \phi^{u_0}_l (\rho )
\end{equation}
with
\begin{equation}
\phi^{u_0}_l(\rho)
\mathop{ \rightarrow }_{\rho \to  \infty }
\sin \left( \rho - \frac {1}{2} l\pi + \delta^0_l \right)
\end{equation}
where $\delta^0_l$ is a reference input phase shift.
With the normalization and expansion coefficients so given, 
the complete 
solution functions can be determined 
from Eq.~(\ref{eqn15}) at all $\rho < \rho_0$.  
Thereby one gets the Jost
transformation kernels and thence the sought after potential.

\subsection{The WKB method}

In the WKB approximation with $\lambda = l + \frac{1}{2}$,
scattering phase shifts are defined~\cite{C&S,Newton} by
\begin{equation}
\delta(\lambda) = {\pi \over 2} \lambda - kr_0 + \int^\infty_{r_0}
\left[ K_\lambda(r')-k \right]\ dr'
\label{wkbeqn}
\end{equation}
where $K_\lambda(r)$ describes the local momentum through the interaction
region and $r_0$ is the classical turning point. 
Thus the scattering potential is
embedded in $K_\lambda(r')$ and 
inversion amounts to an integral transformation. To
effect such a transformation it is convenient to consider 
the deflection function
\begin{equation}
\Theta(\lambda) = 2 \frac{d\delta(\lambda)}{d\lambda}\ ,
\end{equation}
where now $\lambda$ is taken as the angular momentum variable.
This deflection function
satisfies an
Abel-like equation, found by applying the
Sabatier transformation,
\begin{equation}
\sigma = kr\left[1 - \frac{V_{WKB}(r)}{E}\right]^{\frac{1}{2}}
\end{equation}
to Eq.~(\ref{wkbeqn}). 
One finds
\begin{equation}
\delta(\lambda) = -\frac{1}{2E}\int^\infty_\lambda 
\frac{Q(\sigma)}{\sqrt{\sigma^2 - \lambda^2}} 
\sigma d\sigma\ ,
\label{delta}
\end{equation}
where $Q(\sigma)$ is a quasi-potential defined  by
\begin{equation}
Q(\sigma)=2E \ln \left(\frac{\sigma}{kr}\right)\ .
\end{equation}
The Abel-like integral equation for $\delta (\lambda)$
can be inverted to give
\begin{equation}
Q(\sigma) = \frac{4E}{\pi} 
\frac{1}{\sigma} \frac{d}{d\sigma}
\int^\infty_\sigma
\frac{\delta(\lambda)}{\sqrt{\sigma^2 - \lambda^2}} 
\lambda d\lambda\ ,
\end{equation}
which can be written in terms of the deflection function as
\begin{equation}
Q(\sigma) = \frac{2E}{\pi} \int^\infty_\sigma 
\frac{\Theta(\lambda)}{{\sqrt{\sigma^2 - \lambda^2}}} d\lambda\ .
\end{equation}
Provided there is a one to one mapping of the transcendental equation
\begin{equation}
k r =  \sigma \exp{\left(\frac{Q(\sigma)}{2E}\right)}\ ,
\label{transc}
\end{equation}
and the energy $E$ exceeds that at which `orbiting' occurs, i.e.
\begin{equation}
E > V(r) + \frac{1}{2} r \frac{dV}{dr}\ ,
\end{equation}
then the Sabatier transformation equation provides the relationship
from which the scattering potential can be found, namely
\begin{equation}
V_{WKB}(r) = E\left\{ 1 - \exp{\left[-\frac{Q(\sigma)}{E}\right]}\right\}\ .
\end{equation}

For large $\sigma$, the quasi-potentials
decrease so that with $\sigma \to kr$,
$Q(\sigma)  \to V(r)$. As $\sigma \to 0$ however,
the quasi-potentials diverge and the transforms
then lead to the lower limits
$r \longrightarrow r_0$ (the turning point radius),
$V(r)  \longrightarrow E\ $.
However, in practical cases the validity of the WKB approximation
breaks down at a radius larger than $r_0$,
when the transcendental relationship
between $\sigma$ and $r$
becomes ambiguous.


\section{Specification of sets of phase shifts}

\begin{figure}
\centering\epsfig{figure=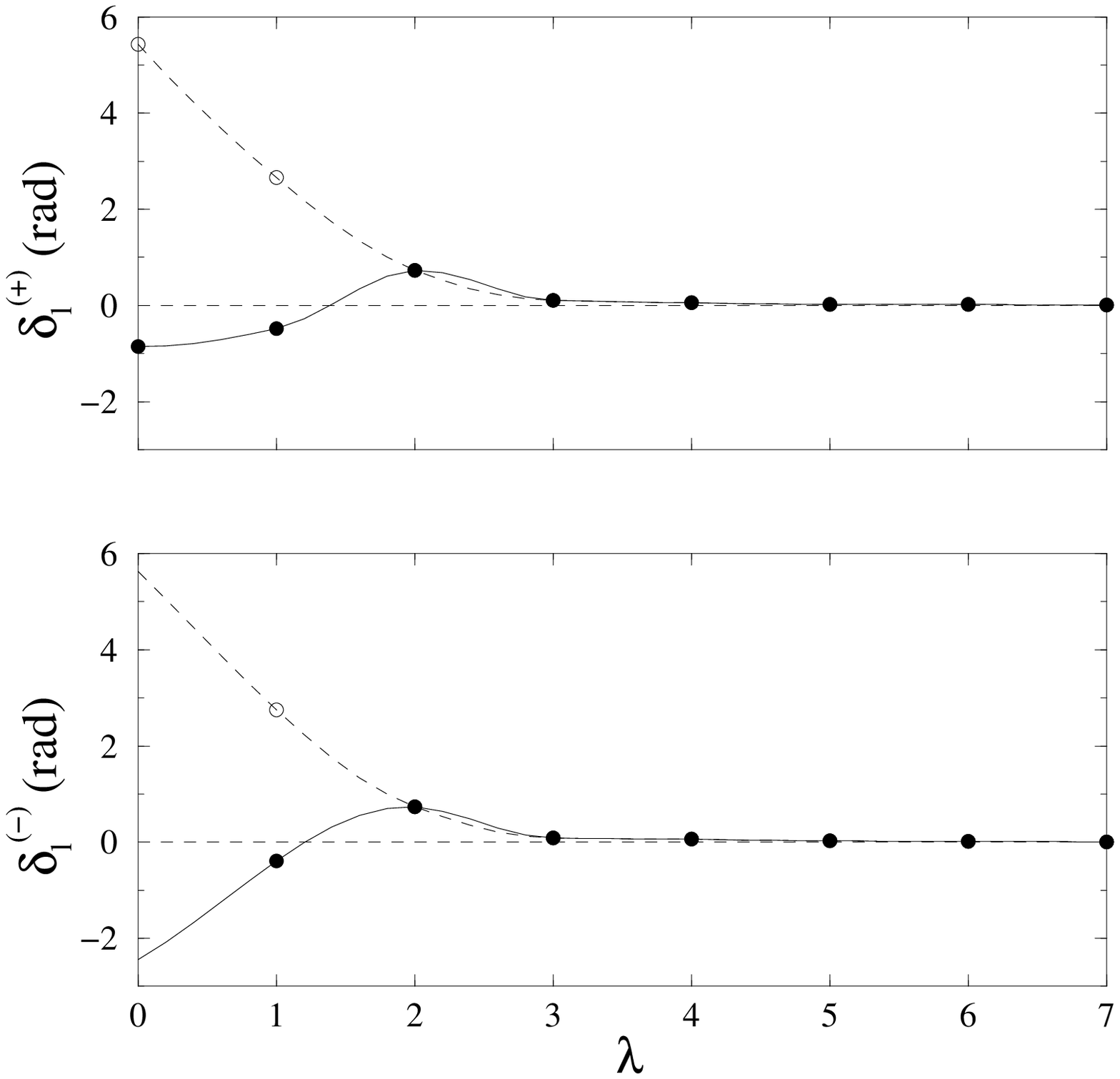,width=\linewidth,clip=}
\caption[]{The scattering phase shifts found from a phase shift analysis
of 5 eV e-Xe scattering data. The filled circles are the results
specified by Gibson {\it et al.}~\cite{Buckman}
while the open circle are the $\pi$-adjusted values.
The results 
of interpolations of the basic two  
sets of (physical) phase shifts
are portrayed by the 
solid and dashed curves.
}
\label{splinecomp}
\end{figure}

The 5 eV e-Xe phase shifts determined by 
Gibson $\it et.al.$~\cite{Buckman} have interesting structure
notably that while the $s$- and $p$-wave phase shifts are negative, 
for all other $l$-values they are positive.
The filled circles in
Fig.~\ref{splinecomp} depict the the phase shifts 
that have been extracted
from the data.
The top most graph identifies the phase shifts 
associated with the $j = l + \frac{1}{2}$ angular momentum
set
while the bottom panel contains those associated with
$j = l - \frac{1}{2}$. It is evident from the data  displayed in this 
figure that the there is only a small difference
between the $\delta^\pm_l$ sets. 
The largest difference occurs with  the $p$-wave
phase shifts, and that is only of order $0.1^c$.

As the phase shift analyses of the e-Xe scattering 
data gave negative values for the 
$s$- and $p$-wave phase shifts,
one can expect~\cite{Newton}
 scattering potentials that have a short ranged repulsion.
But for e-Xe scattering it is known that the
potential should be
attractive at all radii and
especially so near the origin where the 
incoming electron should feel essentially only the
presence of the nucleus. Then one would expect
the phase shifts for low $l$-values to
be positive. 
Such can be formed with the phase shift
values having a monotonic decrease with $l$
by the addition of $2\pi$ to $\delta^{(+)}_0$ and $\pi$ to 
$\delta^\pm_1$. 
We define such modulated values
as the $\pi$-adjusted phase shifts hereafter.
Naturally multiples of any integer
amount may be used, 
but this new set is the simplest.  The 
new ($\pi$-adjusted) values are shown 
by the open circles in Fig.~\ref{splinecomp}.
Associated with such phase shifts are purely 
attractive interactions which are expected in the
physical potentials for electron-atom scattering.

To investigate 
the effect of additional input
on the form of the inversion potentials,
the two data sets were interpolated.
Several interpolations were made  
seeking suitable input for the different
inversion schemes. 
A many point interpolation was made 
on each phase shift set
to obtain the input for the 
WKB inversion scheme.
Values of the 
phase shift functions had to be found at quite small step sizes,
$\Delta l$, since in the WKB we have to evaluate not only
 the deflection functions but also integrate over them
(numerically).  A step size $\Delta l$ of $0.01$ was used.
Also two extended sets of input phase shift
values were generated for use with the 
N-S scheme. One had $\Delta l= 0.5$ and the
other $\Delta l= 0.2$. This was done to 
assess the effect of differing numbers
of non-physical input on the N-S inversion potentials.
The sets of interpolated phase shifts obtained using  
$\Delta l= 0.2$ are displayed in Fig.~\ref{splinecomp},
with the solid and dashed curves giving the 
results of those (spline) interpolations.
Clearly the phase shift functions
so specified  are no longer equivalent and so
we expect any inversion process that requires such 
functions as input to give different
inversion potentials.


\section{Results and discussion}

The results we have obtained using the N-S inverse scattering theory
are discussed first and, subsequently, those found from our
WKB study of the chosen two phase shift functions 
are considered.  We present 
in three subsections,
the potentials that result,
the phase shifts found 
from solutions of
the Schr\"odinger equations
containing those potentials, 
and the 
cross sections that ensue in each case.

\subsection{The results from N-S inverse scattering theory}

Our N-S studies have lead to
six inversion potentials;
three found
by using 
the original phase shift values of Gibson {\it et al.}~\cite{Buckman}
and the other three obtained by using the $\pi$-adjusted
phase shifts.
For each case, we first calculated
the N-S inversion potentials using
as input solely the 
phase shifts corresponding to the physical $l$ angular
momentum set $l\in {0, 1, 2,..., 7}$.  
Such results we identify as 
case 1 results, e.g. case 1 inversion potentials from the 
$\pi$-adjusted phase shift sets.
Two other calculations have been made
with N-S inversion. First the N-S inverse scattering theory equations
have been solved using the discretization
$\Delta l = 0.5$;
the results we identify by the designation case 2.
The third set of N-S calculations were made
using $\Delta l = 0.2$
to give what we term case 3 results.

\subsubsection{The N-S inversion potentials}

Physical arguments dictate that the e-Xe scattering potential 
for 5 eV electrons  should be real 
(the energy is below the first threshold) and 
to be attractive, with long range behavior
of ${-|\alpha |/ r^4}$ and a short ranged one of 
${-Ze^2/ r}$. One may also expect
that the intermediate range potential would be essentially a 
monotonic function between the extremes
as the charge density of the atom is believed to be
a smooth function

The potentials resulting from the inversion of the phase 
shifts based upon the original 
set~\cite{Buckman}
are all strongly repulsive at small radii
and hence are not considered physically significant. 
They also have marked oscillation in both their central
and spin-orbit results.
When used in the Schr\"odinger equations however,
the solutions do reproduce the input phase shifts 
quite well; comparable to the results we show
subsequently.
But as the inversion potentials are not
consistent with the form of e-Xe potential
dictated by knowledge of that scattering system,
we consider those inversion results no further.

\begin{figure}
\centering\epsfig{figure=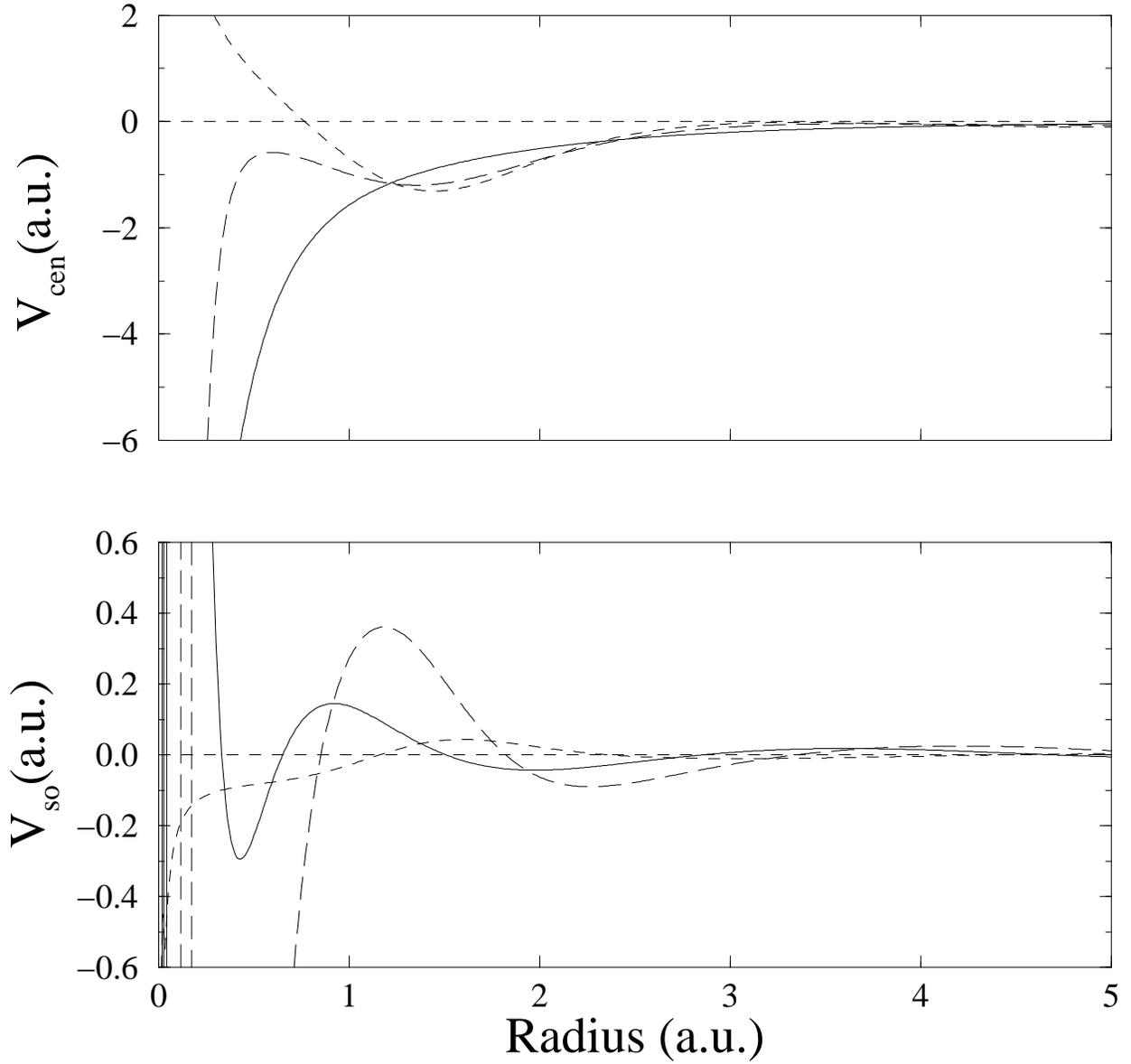,width=\linewidth,clip=}
\caption[]{Potentials (central top; spin-orbit bottom)
 obtained from N-S inversion
of the three sets of phase shift values described in
the text and formed by using the $\pi$-adjusted 
phase shift values.
The dashed, long dashed
and solid curves depict the results
designated cases 1, 2, and 3 in the text respectively.}
\label{NSPOTS}
\end{figure}

In 
Fig.~\ref{NSPOTS} 
the potentials obtained by inversion of
the $\pi$-adjusted phase shift values
are displayed.
The top and bottom segments portray
the central and spin-orbit components of the potentials
respectively.
The dashed, 
long dashed, and solid curves
represent the case 1, case 2, and case 3 potentials
respectively. 
The dashed curves in this figure are identical
to the results of inversion found using the
original phase shift values of Gibson {\it et al.}~\cite{Buckman}
That is as should be since
within the N-S inversion scheme made with just the 
phase shift values specified at the physical angular momenta,
modulo $\pi$ adjustment means that one uses
exactly the same expansion wave functions in defining the
internal matrices.
But the other cases have quite different 
outcomes.
The potentials shown in Fig.~\ref{NSPOTS} 
tend to the physical expectation and clearly
demonstrate that the inclusion of greater numbers of phase shifts at
non-integer angular momentum lead to
smoother, more realistic potential forms. 
The  inversion potential found using
the phase shifts at solely the physical
angular momenta does not represent a structure 
expected for 5 eV electrons on Xe atoms as it 
has the
short ranged repulsion.
However between
0.75 and 2.5 a.u., that (case 1 central) potential
has an attractive well with a depth of -1.3 a.u. 
while  beyond 2.5 a.u. it
behaves approximately as
$ r^{-4}$. 
The spin-orbit
component of the case 1 potential is very small, is weakly 
attractive in the vicinity of  1 a.u., mildly
repulsive between 1 and 2.5 a.u., and after 
that it is essentially zero.

Obviously the most realistic potential comes with the case 3
potentials found using the $\pi$-adjusted phase shift
sets.
That concurs with the
hypothesis that the 
inverse scattering theory result stabilizes with increase in the
 number of non-integer angular momenta
phase shifts commensurate with 
numerical accuracy of evaluation.
Essentially the angular momentum step size should be small in
comparison to the number of significant partial wave
input data ($l_{max}$).
 This case 3 potential, shown in Fig.~\ref{NSPOTS}
by the solid lines, has exactly the structure 
one would associate with e-Xe scattering. 
At small radii it is strongly attractive and of $r^{-1}$ form,
with a smooth transition to a long range
$r^{-4}$ tail. 
There is a smooth transition between those regions.
The spin-orbit 
results also are more reasonable.
The case 3 (with $\pi$-adjusted
phase shifts as input),
 spin-orbit potential
is not as extensive as 
the others.
But the spin-orbit potentials
are all small in general (save for the 
naturally occurring divergence at the 
origin)
 and so these three results
do not by themselves indicate convergence. 
The reproduction of phase shifts
and observables, however, tend to
suggest that 
the results we show are reasonable.

\subsubsection{Reproduction of the phase shifts}

An indication of the success of an inversion scheme is
to reproduce  the input phase shifts
from solutions of the Schr\"odinger equations using
the inversion potentials.
In this case such reproduction is reasonably good but not exact;  
perhaps being  a measure
of the LHF approximation.

\begin{figure}
\centering\epsfig{figure=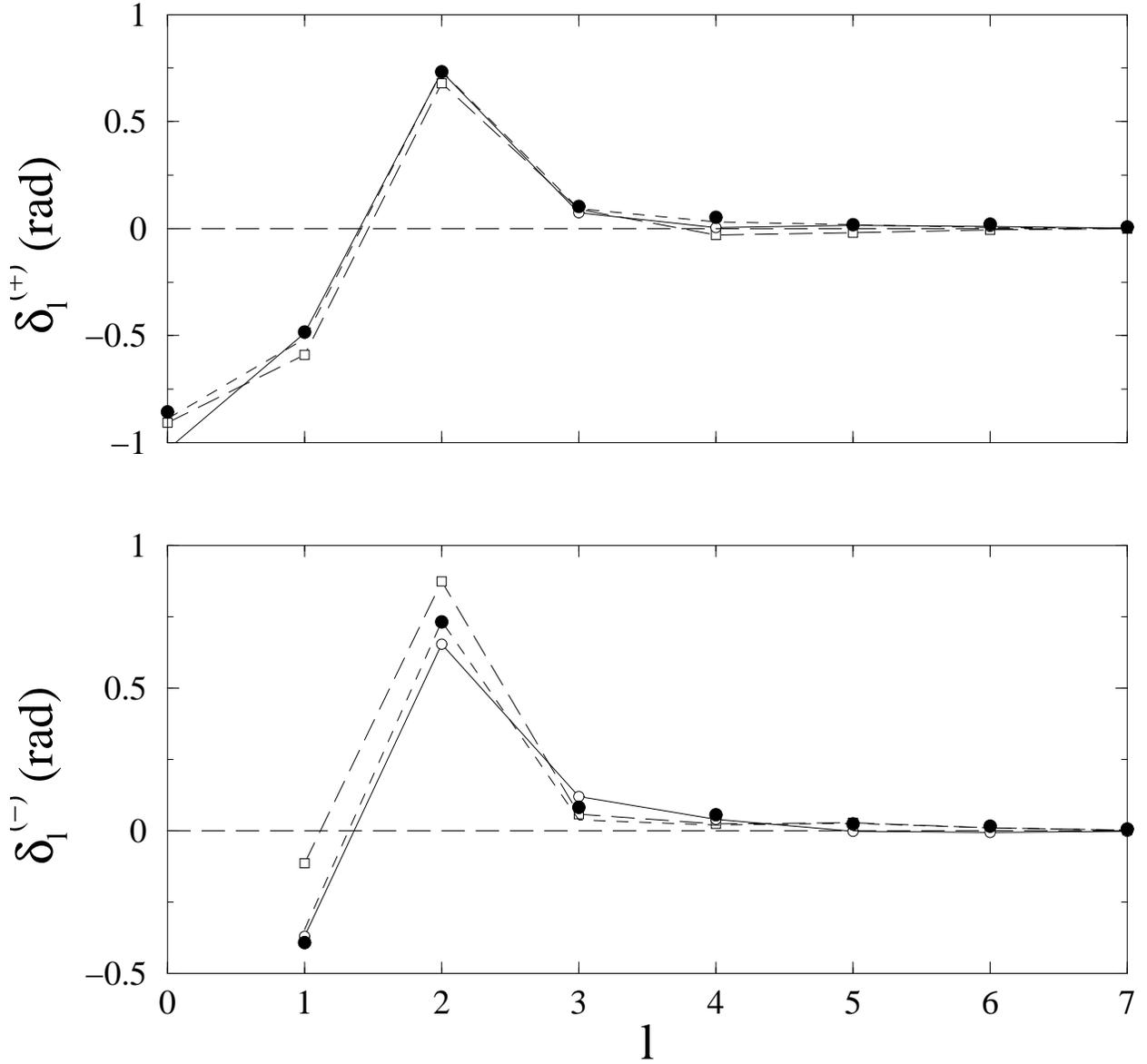,width=\linewidth,clip=}
\caption[]{ 
The 5 eV e-Xe phase shifts obtained from
the potentials shown in Fig.~\ref{NSPOTS} compared with
the values specified by Gibson {\it et al.}~\cite{Buckman}.
Note that the lines are simply to guide the eye
and to identify the three cases. The notation is as used
in Fig.~\ref{NSPOTS}.
}
\label{NSphases}
\end{figure}

In Fig.~\ref{NSphases}, the original
phase shift values~\cite{Buckman} 
are portrayed by the filled circles and
are compared with those
obtained (at integer angular momenta)
using each of the three inversion potentials defined by
the $\pi$-adjusted input sets. 
Case 1 results lie within the filled circles,
case 2 values are shown by the open squares, and case 3
gave the results portrayed by the open circles. 
The three lines now are meant only to guide the eye
by connecting the phase shift values
arising from use of
the three inversion potentials.

The best reproduction of the phase shifts 
defined from experiment~\cite{Buckman}
is found by using the case 1 inversion potential,
notwithstanding that
the potential 
contains unphysical characteristics. 
Essentially, the inversion scheme in this case
produces
a potential that has been defined from 
just the two sets of
eight phase shift values found
by Gibson {\it et al.}
at the physical values of $l=0$ to $l=7$.
Case 2 and 3 potentials, on the other hand, were
built using many more phase shifts specified  at non-integer
$l$-values and as the inversion potentials then 
seek to reproduce all of those extra 
values equally well, small variations in the 
results at the 8 physical $l$-values from the 
sets of Gibson {\it et al.}
can result.
The choice of those 
additional (unphysical) phase shifts then is
crucial if the resulting potentials are to reproduce 
the eight physical values very well.
One has to balance the need for a sufficiently
large basis so that the inversion potential 
has stabilized to the proper (physically credible) limit,
against the numerical accuracy one needs to achieve 
with reproduction of the physical phase shifts and scattering 
data.

A source of possible error in addition is the choice
that must be taken for 
the phase shift at the  unphysical 
point $\delta^{j = -\frac{1}{2}}_0$. That value 
is needed
in the calculations of both $\tilde{\delta}$
and $\hat{\delta}$ and also in the N-S scheme. 
This choice has the potential
to introduce error since its 
value influences the interpolation.
A poor choice of this value can become evident 
when the inversion potential is used to recalculate
the phase shifts, particularly for the low-$l$  partial waves.
A reasonable choice seems to be to set the
phase shift $\delta^{j = -\frac{1}{2}}_0$ equal to 
$\delta^{j = +\frac{1}{2}}_0$.
Admittedly this is
an arbitrary point. In this study, 
two choices  were  considered; the first being to take 
$\delta^{j = -\frac{1}{2}}_0 = \delta^{j=+\frac{1}{2}}_0$ and 
the other to use
an Akima spline to determine the
value by extrapolation. In this study allowing the 
Akima spline to determine this point was slightly more successful.

\subsubsection{The cross section from the N-S inversion potentials}

\begin{figure}
\centering\epsfig{figure=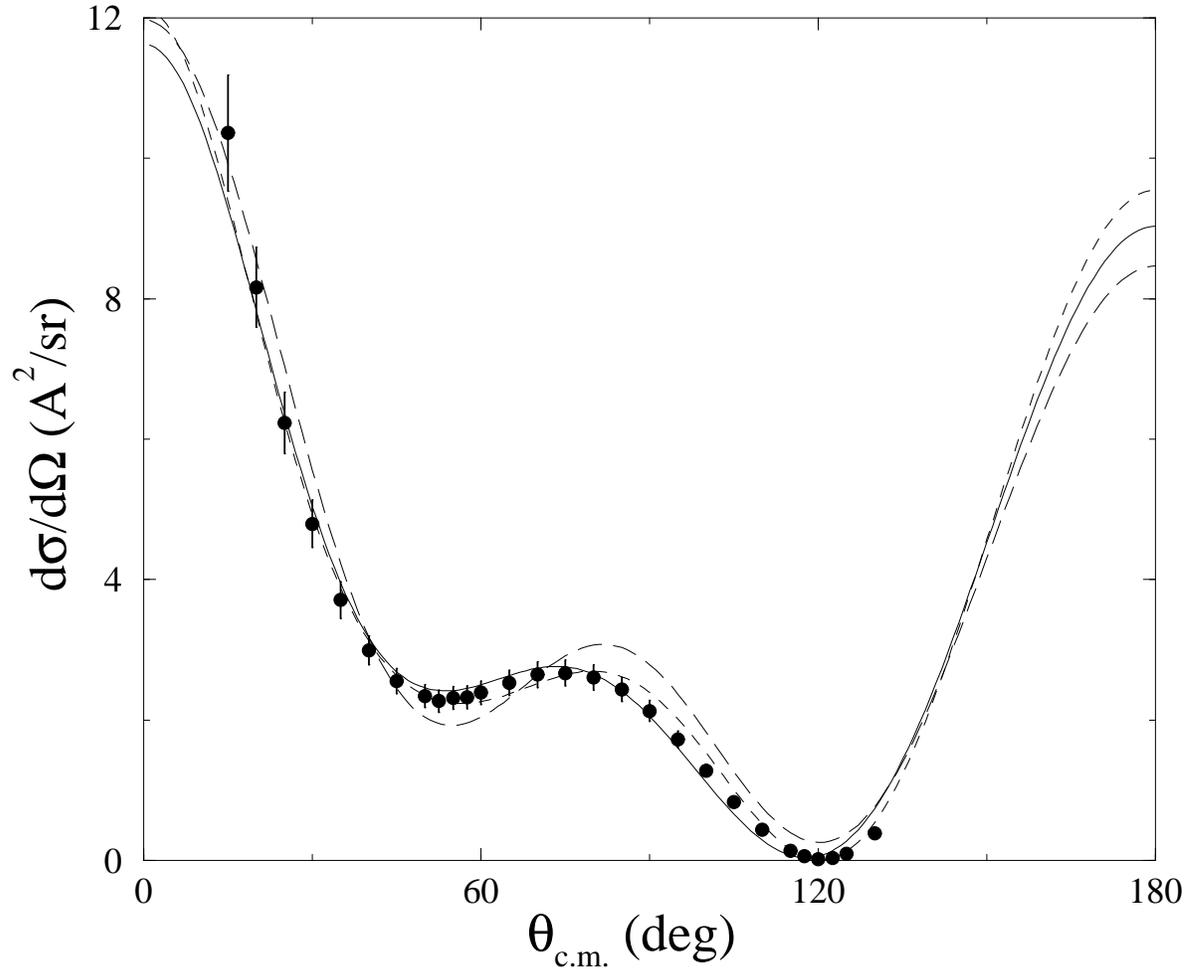,width=\linewidth,clip=}
\caption[]{
The 5 eV e-Xe differential cross sections obtained from
the potentials shown in Fig.~\ref{NSPOTS} compared with
the data of Gibson {\it et al.}~\cite{Buckman} (dots with error bars).
The notation defining the results from the three cases is
as in Fig.~\ref{NSPOTS}.
}
\label{NSxsec}
\end{figure}

Although the potentials all look reasonably good, 
particularly that found from case 3
with the $\pi$-adjusted input phase shifts,
the further test
of the inversion results 
is to see
how accurately use of the potentials
reproduce experimental data.
This is displayed in Fig.~\ref{NSxsec} 
wherein the experimental
data (with error bars) as found by Gibson {\it et al.}~\cite{Buckman}
are compared with 
the cross sections calculated
from the three inversion potentials
obtained using the $\pi$-adjusted phase shift sets.
Those results found using the case 1, case 2, and case 3
 inversion potentials
 are depicted by the
dashed, long dashed , and solid curves
respectively.
Further we display the results here on a linear scale
to distinguish the bulk of the results in relationship to the error
bars.  Later when discussing the WKB calculated cross sections,
we will display also the case 3 N-S comparison with data
shown on a semi-logarithmic plot.  That emphasizes the comparison of 
results with data at the large scattering angles
and particularly in the vicinity of the minima as 120$^\circ$.
As one might expect from the reproduction of the  
phase shifts, we see in Fig.~\ref{NSxsec} that a good reproduction
of the cross-section data is found with the case 1 results.
That result passes through
the error bars of most data points.
The case 2 cross section has
similar structure to the experimental data, 
but the shape is slightly at variance falling just outside
the error bars associated with a number of the data.
The case 3 cross section however is in
 excellent agreement with most of the experimental
data. In general it falls within most of the data error bars 
except at the larger scattering angles; the latter 
indicative of the
phase shifts at integer values of $l$ 
not being reproduced with sufficient accuracy.

\subsection{The results from semi-classical WKB inversion theory}

The WKB inversion results have been obtained by forming 
the $\tilde{\delta}(\lambda)$ and
$\hat{\delta}(\lambda)$ phase shift functions 
and using them to  evaluate two
quasi-potentials.
The potentials that result are discussed in the first 
subsection. Subsequently we present and discuss the phase 
shift reproductions and the cross sections that
result on using those inversion potentials.

\subsubsection{The WKB inversion potentials}

\begin{figure}
\centering\epsfig{figure=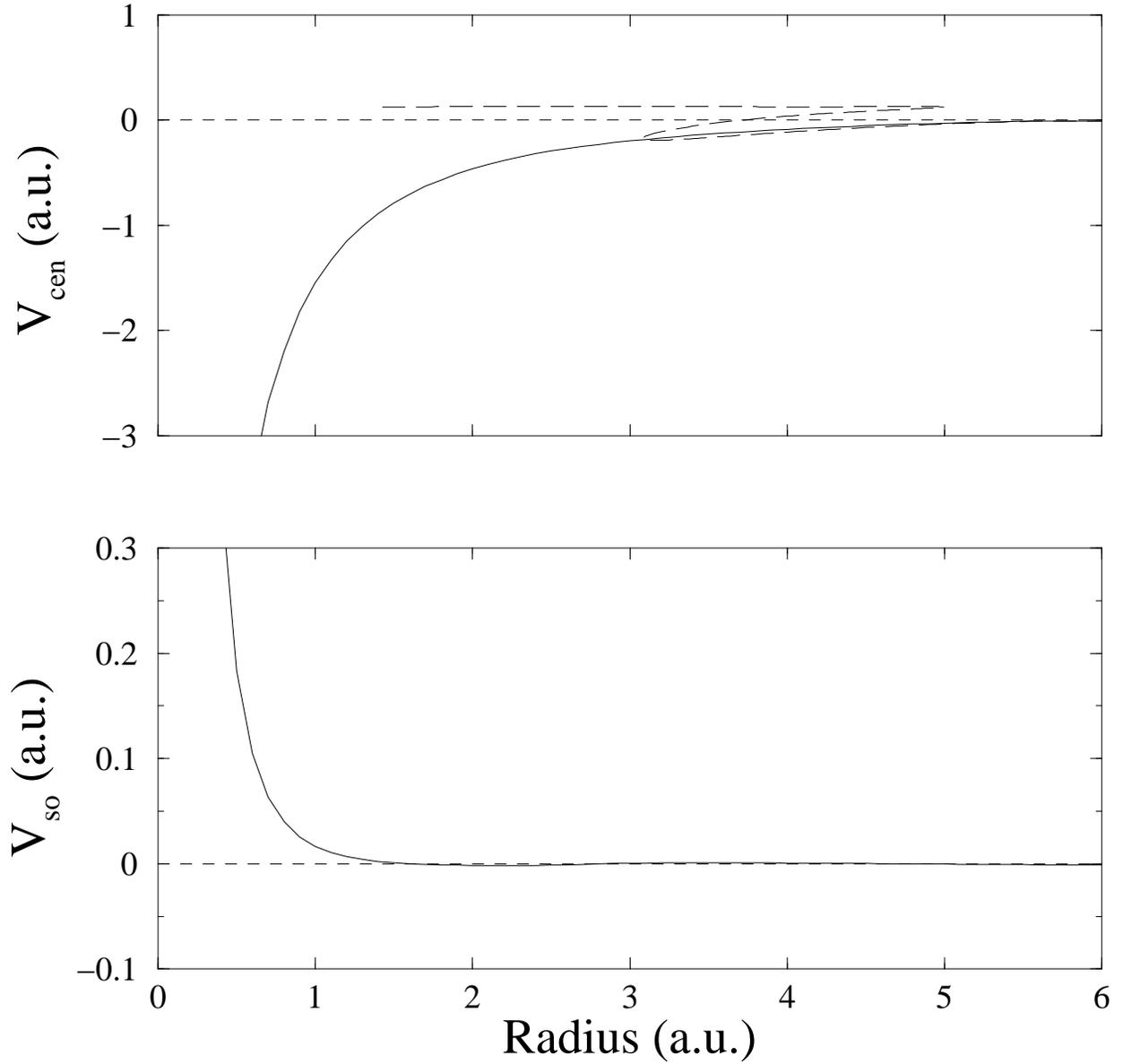,width=\linewidth,clip=}
\caption[]{Potentials (central top; spin-orbit bottom)
 obtained from WKB inversion
of the phase shift functions described in
the text.
The solid curves depict the results
found using the $\pi$-adjusted 
phase shifts to define the phase shift function; the long dashed
curve gives the (central) potential
found using 
actual Gibson {\it et al.}~\cite{Buckman}
tabled values for that purpose.}
\label{WKBV}
\end{figure}

The inversion potentials we have found using the semi-classical
 WKB methods are presented in Fig.~\ref{WKBV}.
Once again the central potential is shown in the
top section of the 
figure; the
spin-orbit potential in the bottom. 
Clearly
two very different
structures have been generated.
The result from using the $\pi$-adjusted functions 
has physically sensible characteristics but
that found using the 
deflection function defined from the original phase shifts
does not. Indeed the inversion procedure based upon the original 
phase shift set does not lead to a
spin-orbit potential one can identify as anything sensible
and so that is not displayed.

The $\pi$-adjusted phase shift data sets give smooth monotonic 
phase shift functions
and hence work well with the WKB scheme. Furthermore it leads to
the form of potential (the continuous lines in
Fig.~\ref{WKBV}) one would expect
for the central component of an e-Xe interaction,
i.e.
a smooth function which
is highly attractive
toward the origin
and has an attractive $r^{-4}$ long range behavior.
The spin-orbit component of the potential would also be small and short ranged.
Essentially the potential from this WKB analysis that is displayed in 
Fig.~\ref{WKBV} has
this desired prescription.
However, given that the input was not optimally smooth 
at large angular momenta (close inspection of the input data
showed that there were
small oscillations in the phase shift functions),
there are 
small oscillations at large radii
in the  WKB inversion potential. 
Nevertheless, by smoothing,
the long range WKB inversion  potential does 
behave on average like $r^{-4}$ and 
the central component of this WKB inversion potential 
resembles that found using the N-S
scheme.

\begin{figure}
\centering\epsfig{figure=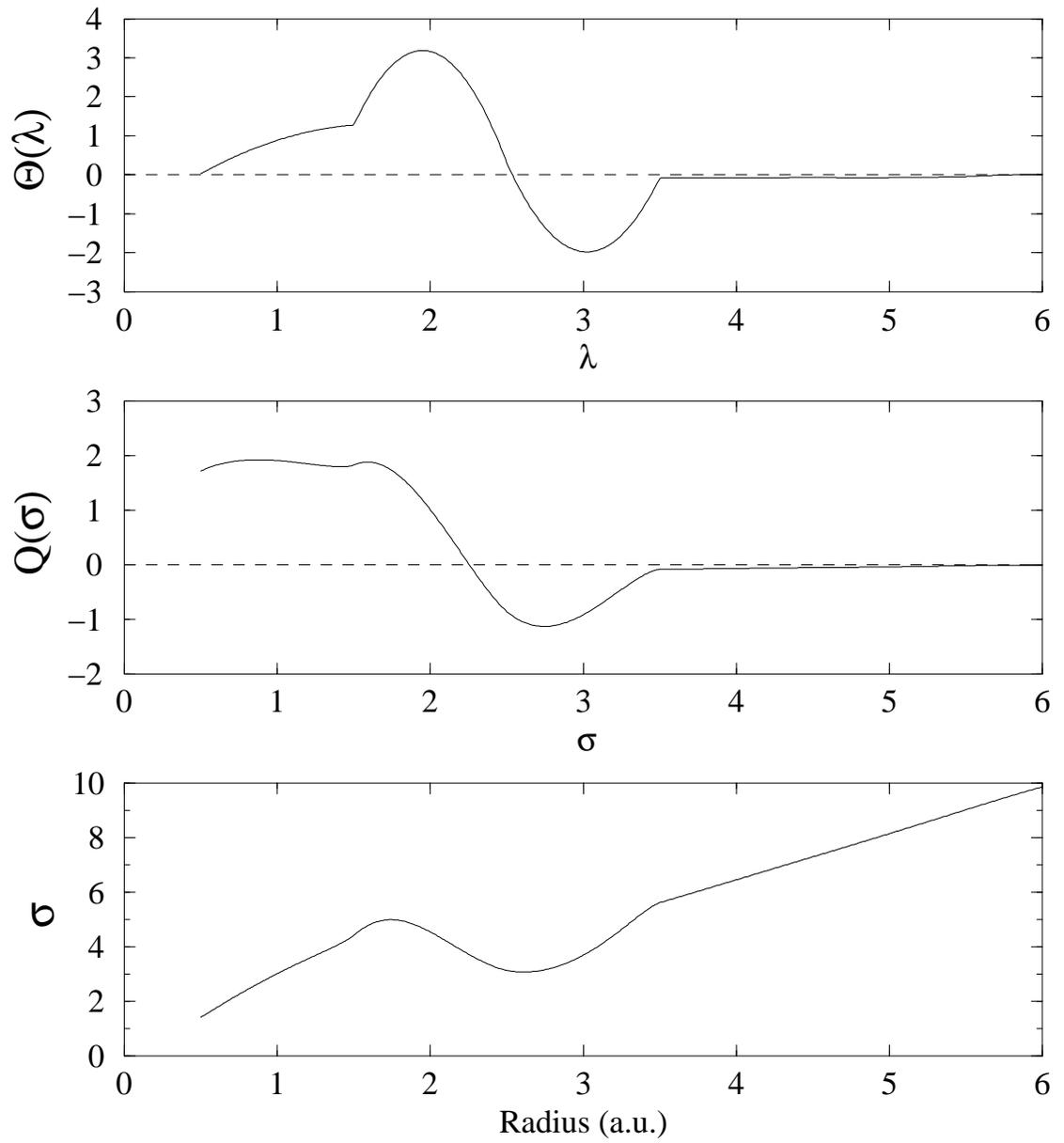,width=\linewidth,clip=}
\caption[]{
The deflection function (top), the quasi-potential (middle),
and the $\sigma$ vs $r$ plot (bottom)
from the WKB study framed upon
the original phase shift values of Gibson {\it et al.}~\cite{Buckman}.
}
\label{defquas}
\end{figure}

Only the central potential found
from WKB inversion of the original 
phase shift data~\cite{Buckman} is shown
in Fig.~\ref{WKBV} 
(by the long dashed  line). 
Quite evidently it is nonsensical.
It exhibits ambiguous behavior at many radii,
resulting from a loss of 1:1 correspondence between $\sigma$
and $r$ in this case.
Apparently if the input phase shift function contains 
a large degree of curvature
then that input is unsuitable for use with the WKB procedure.
This is emphasized when one considers the 
deflection function, quasi-potential, and the 
associated $\sigma$ vs $r$ 
plots.
The deflection function resulting from differentiating the 
phase shift function defined from interpolation
of the original phase shift values, has
a rapid variation as is evident from the top panel in 
Fig.~\ref{defquas}. Consequently the quasi-potential
will also  be quite structured and that is shown in the middle panel of
Fig.~\ref{defquas}. As a result the stability
condition breaks down at a fairly large radius.
 This condition, the correspondence between $\sigma$ and $r$,
is displayed in the bottom section of Fig.~\ref{defquas}. 
Clearly between 3 and 5 a.u. there is an ambiguous
relation with $\sigma$ and so one finds ambiguous values of the
associated 
WKB inversion potential in that region.
Indeed one can only hope
to ascribe a sensible result 
with this input to the WKB inversion
for radii then in excess of 4 - 5 a.u.

\begin{figure}
\centering\epsfig{figure=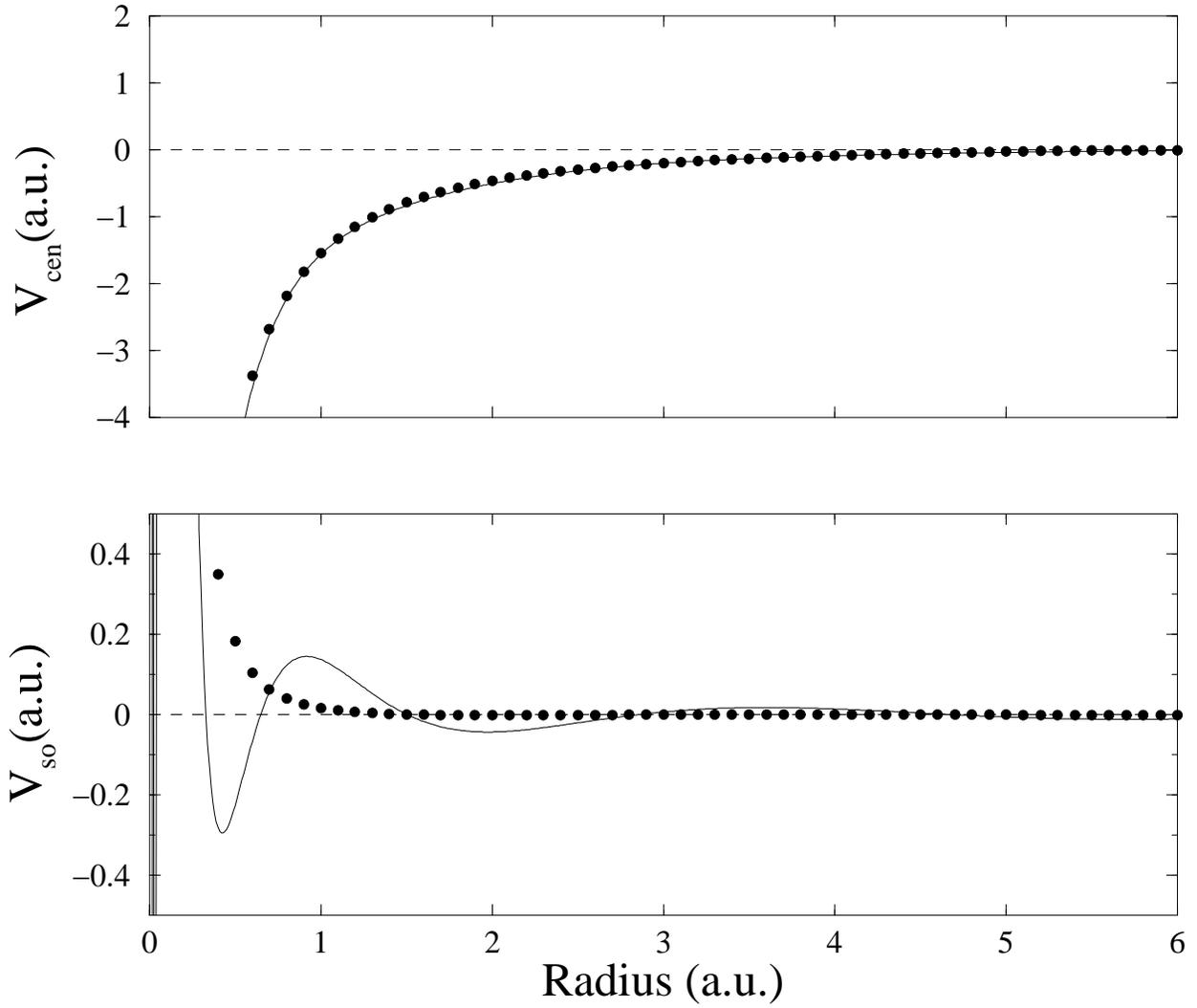,width=\linewidth,clip=}
\caption[]{
Comparison of the WKB inversion potentials (filled circles)
with those of the case 3 N-S inversion study made using
phase shift sets from interpolation of
the $\pi$-adjusted phase shifts of Gibson {\it et al.}~\cite{Buckman}.
}
\label{WKB-NS0.02pots}
\end{figure}

The realistic e-Xe potentials found 
by using the $\pi$-adjusted phase shift sets and
with both  the N-S (case 3)
and WKB inversion schemes, are compared  in
Fig.~\ref{WKB-NS0.02pots}. 
The potentials found using the N-S scheme 
are shown by the solid curves while those found
by WKB inversion are displayed by the filled circles.
The latter are taken to the limit radius
at which the 1:1 correspondence between $\sigma$ and $r$ is maintained.
Overall 
the (semi-classical) WKB potentials 
compare very well with those found using  the (full quantal)
N-S scheme.
Clearly the central potentials look almost exactly the same.
There are differences
between those two which become apparent
when the observables are calculated.
On the other hand,
the spin-orbit WKB potential is very smooth 
and weak in comparison to the 
corresponding N-S
component. Also  it is repulsive at all radii while the N-S potential
has a small attractive well between approximately 0.3 and 0.6 a.u.

\subsubsection{Reproduction of the phase shifts}

\begin{figure}
\centering\epsfig{figure=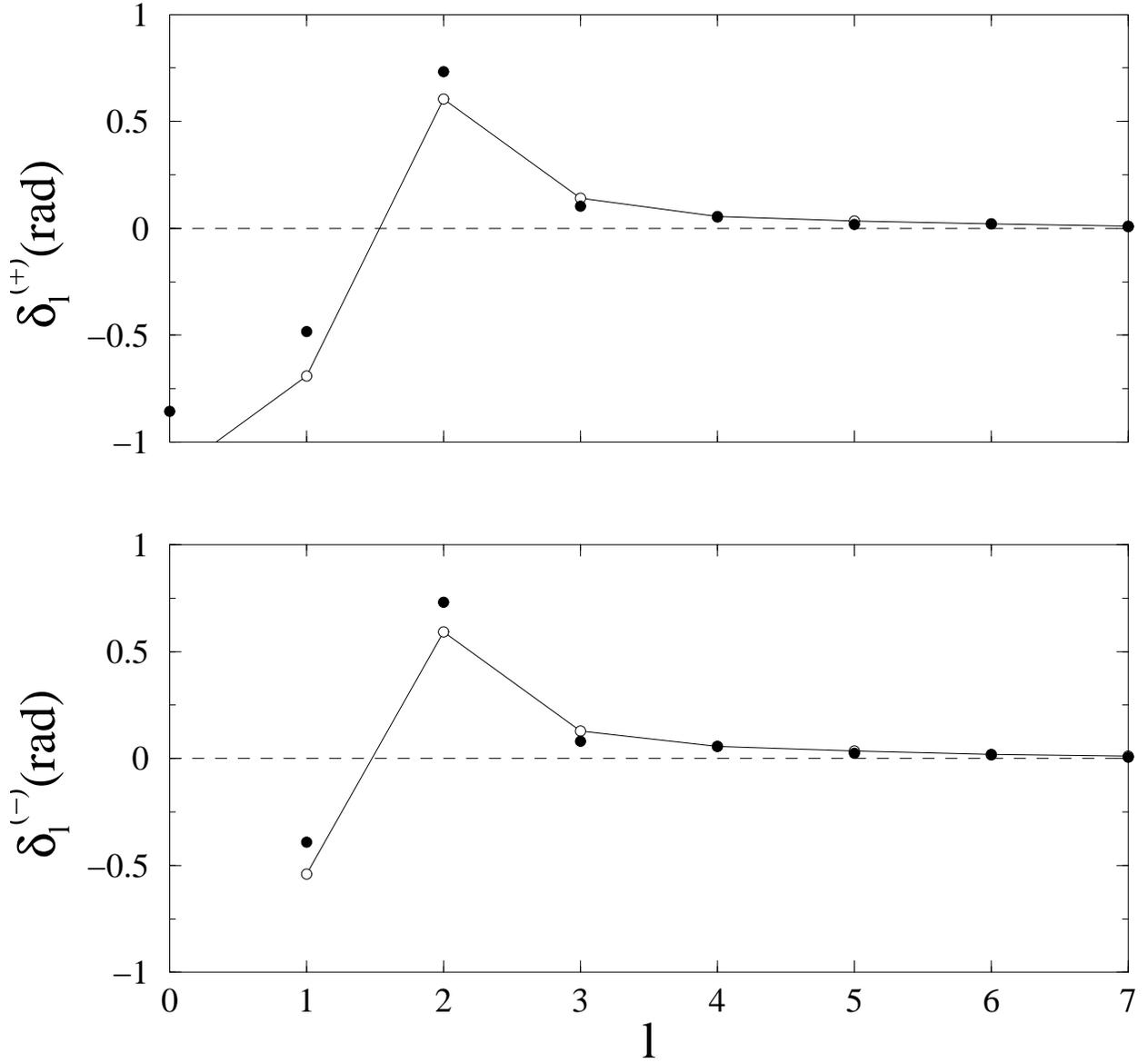,width=\linewidth,clip=}
\caption[]{
The 5 eV e-Xe phase shifts obtained from
the (realistic) potential shown in Fig.~\ref{WKBV} compared with
the values specified by Gibson {\it et al.}~\cite{Buckman}.
Note that the lines are simply to guide the eye.
}
\label{e-Xephas}
\end{figure}

Despite the very pleasing form generated for the potential using 
the WKB scheme, the reproduction of the
phase shifts, although reasonable, 
is not as good as those generated
using the N-S scheme.  However this discrepancy
is not disconcerting given that 
 the WKB scheme is a semi-classical approximation.
The 
phase shifts found from our WKB calculation
are shown by the open circles 
in Fig.~\ref{e-Xephas}, and joining lines have been 
included solely to guide the eye.
The data of Gibson {\it et al.}~\cite{Buckman}
again  are represented by the
filled black circles. As with the N-S case, part of the discrepancy 
between the calculated values and the data may
be due to the ambiguity with the interpolation required
to specify the phase shift functions
and particularly for
points below $l = 1$ for the $j=l - \frac{1}{2}$ input set.

A spline was used to determine the deflection function, 
and that influences the quasi-potential and ultimately also 
the potential.
Given that points found from the  $j = l - \frac{1}{2}$ 
input set 
are extrapolated for values of $\lambda < \frac{1}{2}$, 
ambiguity in those values is inevitable.
That ambiguity persists
when
these splined values are used to determine the input functions
$\tilde{\delta_l}(\lambda)$ and $\hat{\delta_l}(\lambda)$.
That may result in a 
poor reproduction of the physical phase shifts,
particularly of the $s$- and $p$-wave values when the
solutions of the Schr\"{o}dinger equations 
specified with the relevant 
WKB potentials are 
found 
to complete the study loop.

\subsubsection{The cross section from the WKB inversion potentials}

\begin{figure}
\centering\epsfig{figure=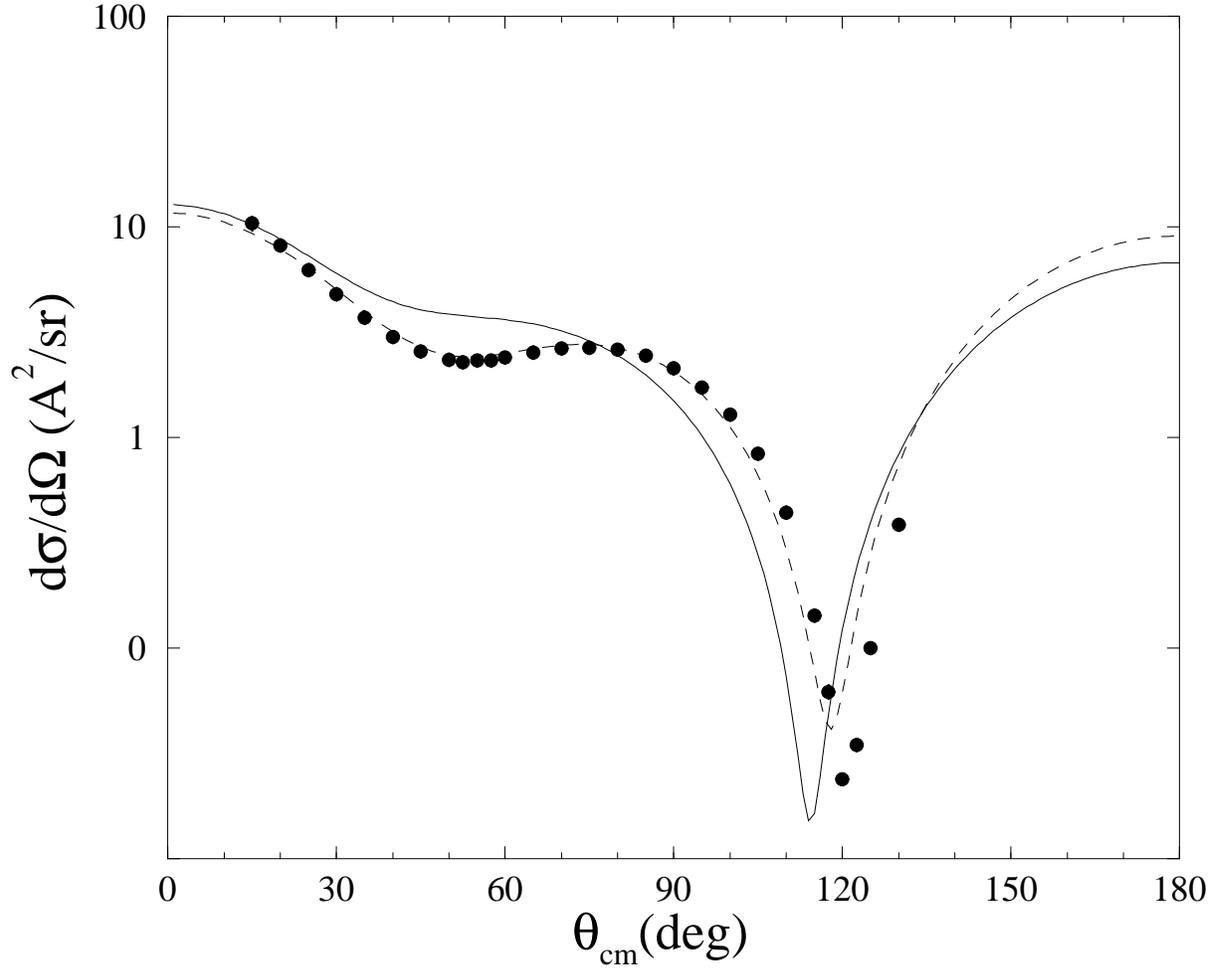,width=\linewidth,clip=}
\caption[]{
The 5 eV e-Xe differential cross section obtained from
the WKB potential (solid curve) compared with
the data of Gibson {\it et al.}~\cite{Buckman}
as well as with the case 3 N-S result (dashed curve).
Both inversion studies used interpolations
of the $\pi$-adjusted set
of phase shifts as input data.
}
\label{WKBe-Xexsec}
\end{figure}

Already, from the reproduction of the phase shifts, we suspect that 
with the WKB method, reproduction of the
cross section will not be as good as 
that obtained using the relevant N-S inversion
potentials.
This is indeed true. The
solid curve shown  in
Fig.~\ref{WKBe-Xexsec} depicts the WKB 
cross section result which is compared therein against
the Gibson {\it et al.} data~\cite{Buckman}
and also against the cross section determined by using
the preferred case 3  N-S inversion potential (shown by the dashed curve).
We now present the results in a semi-logarithmic plot
since 
WKB cross section is similar in magnitude to the
data, however it does not display quite the right structure
at any scattering angle. 
The reproduction simply is not as good as that found using the N-S scheme.
This figure  also emphasizes
the mismatch of the case 3 N-S inversion
results at large angles that we commented on earlier.


\section{Conclusions}

Inversion potentials for the 5 eV electron-Xenon atom
 interaction have been  found using
both the full quantal N-S and the semi-classical WKB
inverse scattering theories.
The results
from application of the N-S inversion  were very good.
 Several inputs were used in this
approach, each containing a different number of phase shift
values. 
From solutions of the Schr\"odinger equations specified with
each of the
potentials, `inversion' phase shifts were extracted that
reproduced well the starting 
(physical) phase shift values
and also the empirical cross section.
However, when the input was taken solely to be 
the (physical) phase shifts at
integer-$l$ values of angular  momenta,
the inversion  
potential contains
a marked  repulsion at small radii.
That is not physical.
 With increased numbers of 
input data specified at
non-integer values of angular momentum, 
inversion produced a potential 
with sensible (physically credible) characteristics.

As two disparate inversion methods find central 
e-Xe potentials that are essentially the same and
have pertinent physical properties
of the scattering system,
we are confident 
that the N-S approach has (nearly) converged
and that the central potential obtained by expanded 
$\pi$-adjusted physical phase shift sets
is the appropriate candidate for the  local 
Schr\"odinger interaction.
The spin-orbit potential is reasonable
but more detailed investigations are needed before the 
characteristics found for it can be adopted
with confidence.

The similarity between the
WKB and N-S inversion 
potentials also implies that
the introduction
of unphysical phase shifts in the N-S calculation is essential
if a stable inversion
potential is to be obtained in this case.
That would be so with other energies measured~\cite{Buckman}
and, by implication, for any such electron-atom scattering
at eV energies.
There is the
difficulty however, of accurately specifying the phase shifts
 for non-integer angular momenta. Obviously some kind of
{\it a priori} information regarding the colliding system
is necessary.
In this instance simply $\pi$-adjusting the given phase shift data
sufficed.


%

\end{document}